\documentclass[12pt]{iopart}
\usepackage{graphicx}
\usepackage[sorting=none]{biblatex}
\usepackage[hidelinks,urlcolor=blue]{hyperref}
\usepackage{caption}
\usepackage{subcaption}

\newcommand{\Msol}{\mathrm{M_\odot}}

\begin{document}

\title[]{Waves in a Forest: A Random Forest Classifier to Distinguish between Gravitational Waves and Detector Glitches}

\author{Neev Shah$^{1,2}$, Alan M. Knee$^2$, Jess McIver$^2$, David C. Stenning$^3$}

\address{$^1$ Department of Physics, Indian Institute of Science Education and Research Pune, Maharashtra, 411008, India}
\address{$^2$ Department of Physics and Astronomy, University of British Columbia, Vancouver,
British Columbia, V6T1Z4, Canada}
\address{$^3$ Department of Statistics \& Actuarial Science, Simon Fraser University, Burnaby, BC, Canada}
\ead{neev.shah@students.iiserpune.ac.in}
\vspace{10pt}
\begin{indented}
\item[]
\end{indented}

\begin{abstract}
The LIGO-Virgo-KAGRA (LVK) network of gravitational-wave (GW) detectors have observed many tens of compact binary mergers to date. Transient, non-Gaussian noise excursions, known as ``glitches'', can impact signal detection in various ways. They can imitate true signals as well as reduce the confidence of real signals. In this work, we introduce a novel statistical tool to distinguish astrophysical signals from glitches, using their inferred source parameter posterior distributions as a feature set. By modelling both simulated GW signals and real detector glitches with a gravitational waveform model, we obtain a diverse set of posteriors which are used to train a random forest classifier. We show that random forests can identify differences in the posterior distributions for signals and glitches, aggregating these differences to tell apart signals from common glitch types with high accuracy of over 93\%. We conclude with a discussion on the regions of parameter space where the classifier is prone to making misclassifications, and the different ways of implementing this tool into LVK analysis pipelines. 
\end{abstract}

%
%
%
%
%


\section{Introduction}
Since the first direct detection of gravitational waves (GWs) in 2015, the LIGO-Virgo-KAGRA (LVK) network of detectors have detected many tens of candidates from compact binary coalescences (CBC) over three observing runs \cite{LIGOScientific:2018mvr,LIGOScientific:2020ibl,LIGOScientific:2021djp}. Since gravitational waves from astrophysical sources are very weak, the detectors need a high degree of sensitivity to achieve the required signal fidelity. This also makes the detectors extremely sensitive to terrestrial noise sources, which often appear in the data, hampering searches for astrophysical signals \cite{LIGOScientific:2016gtq,Nuttall:2015dqa}. ``Glitches'' are a broad class of short-duration noise excursions which commonly appear in GW detector data. They are transient, non-Gaussian noise of environmental origin that have persisted throughout all three observing runs thus far. As many glitches have a similar time-frequency morphology to GW transients, they can trigger false alarms in low-latency search pipelines \cite{LIGOScientific:2017tza}. This can lead to wasted time and resources if telescopes attempt to follow-up these false alarms, as well as decrease the statistical significance of real events \cite{LIGOScientific:2017tza}. 

Glitches can exhibit a range of morphologies, and their origin is not necessarily known. In cases where the cause has been identified, the associated glitches can be eliminated \cite{LIGOScientific:2016gtq,LIGO:2021ppb,Virgo:2022fxr}. Despite these efforts, a large number of glitches still persist in the detector data \cite{LIGOScientific:2018mvr,LIGOScientific:2020ibl,LIGOScientific:2021djp}. There have been various software approaches to identify likely glitches and mitigate them by down-weighting the significance of coincident candidates. For example, there are algorithms that use glitches in auxiliary witness channels to predict the likelihood of a glitch in strain data \cite{Essick:2020qpo,Godwin:2020weu}. Such algorithms can also be used to veto glitches entirely, and has been done to exclude candidates identified in the LIGO-Hanford data in previous observing runs, when they occurred at the same time as when snow plows were active on-site \cite{Davis_2021}. Other algorithms estimate the likelihood of whether the event is astrophysical or a terrestrial noise based on estimates of the likelihood of noise producing a similar candidate \cite{Chatterjee:2019avs}, or using coherence information that requires multiple detectors \cite{Cabero:2020eik,Isi:2018vst,Ashton:2019wvo,Pratten:2020ruz}. Transient GW searches will often have additional glitch-exclusion features, such as the excision of loud, short-duration excursions in the data, or \textit{gating} \cite{sachdev2019gstlal,Aubin:2020goo,chu2021spiir,2021SoftX..1400678D,2021ApJ...923..254D}, which are less effective for low SNR glitches.

Current glitch mitigation methods for CBC searches, e.g. chi-squared re-weighting of the signal-to-noise ratio (SNR) are based on the similarity of the morphological fit between the data and the template \cite{PhysRevD.71.062001}. But such methods do not robustly exclude glitches which have similar time-frequency properties to expected GW signals. It has been seen that light scattering can be mistaken for low-mass CBC sources, as well as tomte glitches being mistaken for high-mass CBC sources \cite{LIGO:2021ppb,jarov2023}. Some glitch-vs-signal classification algorithms, including GravitySpy \cite{2017CQGra..34f4003Z,Glanzer:2022avx,Soni:2021cjy} and GSpyNetTree \cite{Alvarez-Lopez:2023dmv} employ time-frequency spectrograms (a variant of which is called a qscan) \cite{Chatterji:2004qg} and image classification via a convolutional neural network to distinguish between different glitch and signal classes. These algorithms perform well; however, if a single detector captures an event candidate, time-frequency morphology alone may not be enough to confidently distinguish between a true signal and a detector glitch, or a signal coincident with a detector glitch. Choudhary et al. \cite{PhysRevD.107.024030} constructed a deep-learning neural network to distinguish between binary blackhole mergers and blip glitches. There has also been work done that aims to infer the astrophysical population of GW sources, even when some noise transients might be included in the inference \cite{Heinzel:2023vkq,2019MNRAS.484.4008G,PhysRevD.102.083026,PhysRevD.102.123022}.

\begin{figure}
     \centering
     \begin{subfigure}[b]{0.4\textwidth}
         \centering
         \includegraphics[width=\columnwidth]{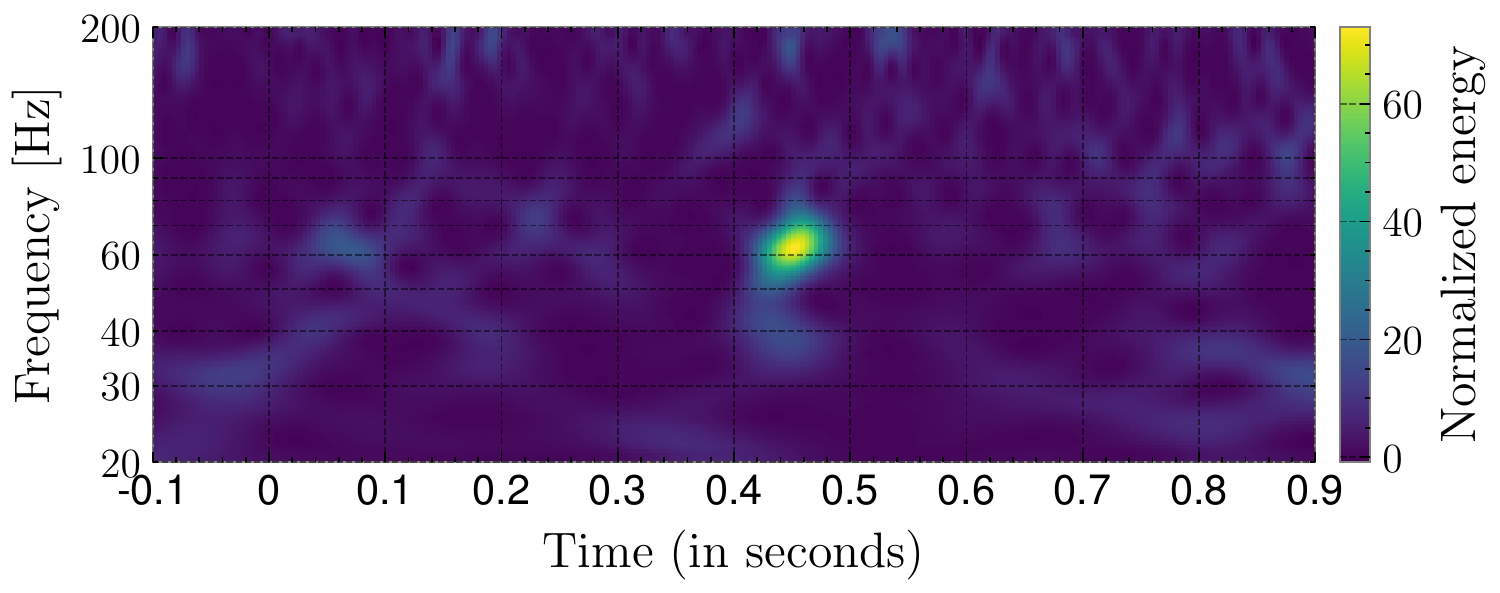}
         \caption{GW190521}
         \label{fig:gw190521}
     \end{subfigure}
     \hfill
     \begin{subfigure}[b]{0.4\textwidth}
         \centering
         \includegraphics[width=\columnwidth]{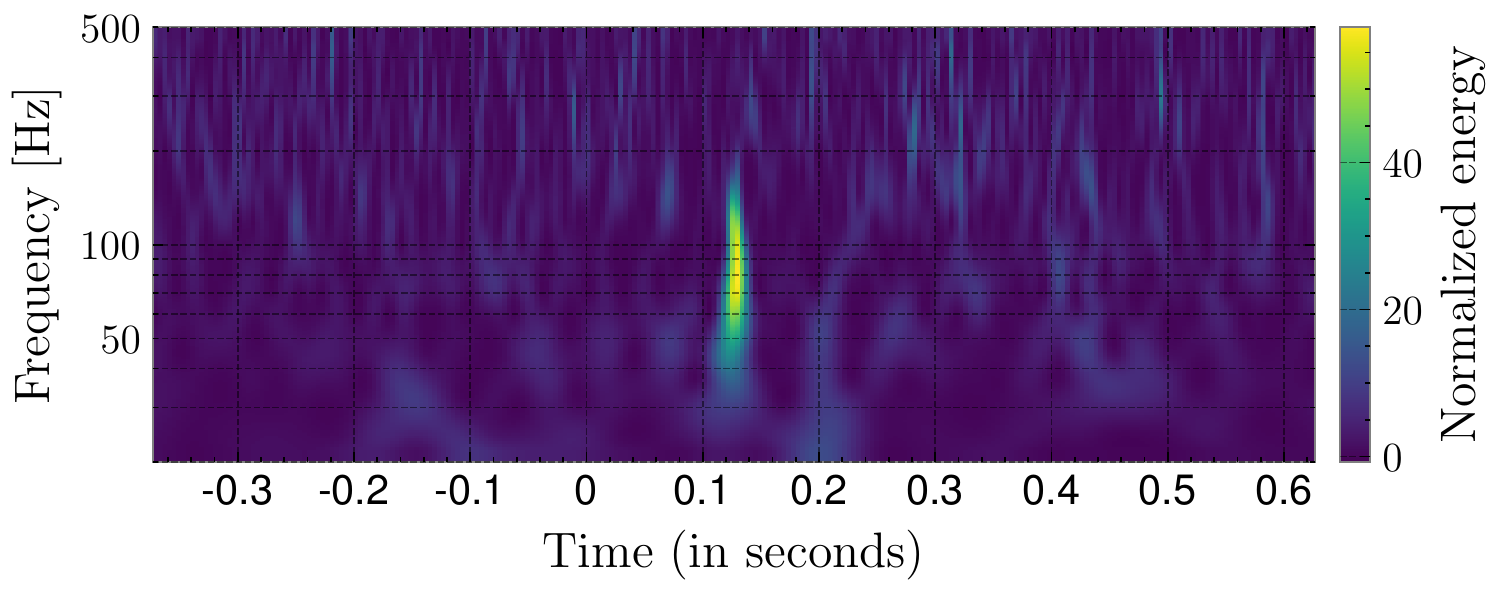}
         \caption{Blip}
         \label{fig:blip}
     \end{subfigure}
          \hfill
     \begin{subfigure}[b]{0.4\textwidth}
         \centering
         \includegraphics[width=\columnwidth]{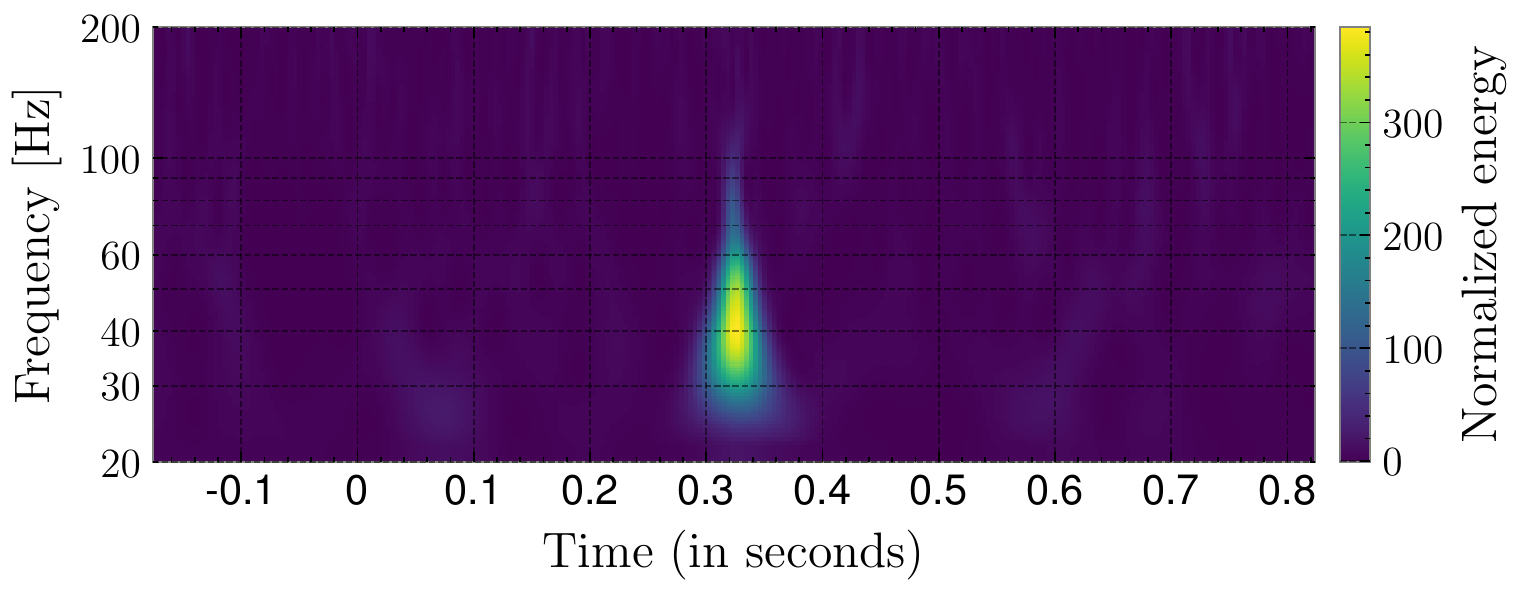}
         \caption{Tomte}
         \label{fig:tomte}
     \end{subfigure}
     \hfill
     \begin{subfigure}[b]{0.4\textwidth}
         \centering
         \includegraphics[width=\columnwidth]{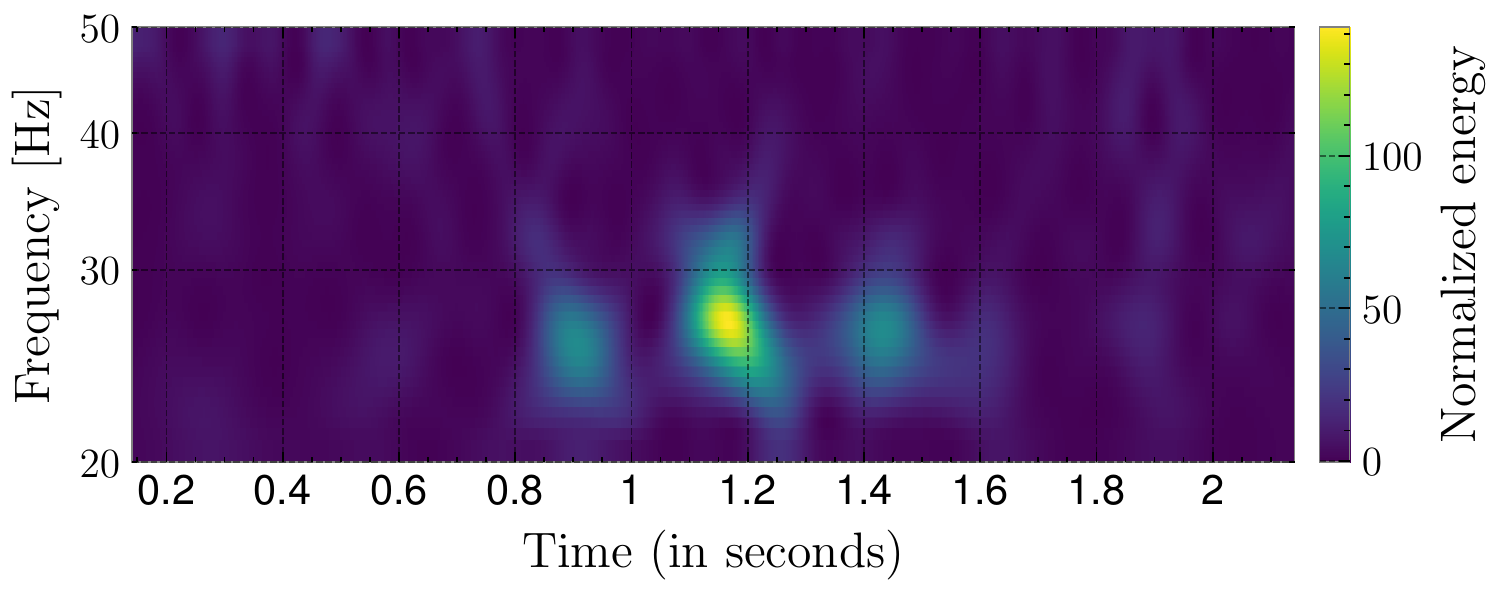}
         \caption{Fast Scattering}
         \label{fig:fastscattering}
     \end{subfigure}
     \hfill
     \begin{subfigure}[b]{0.4\textwidth}
         \centering
         \includegraphics[width=\columnwidth]{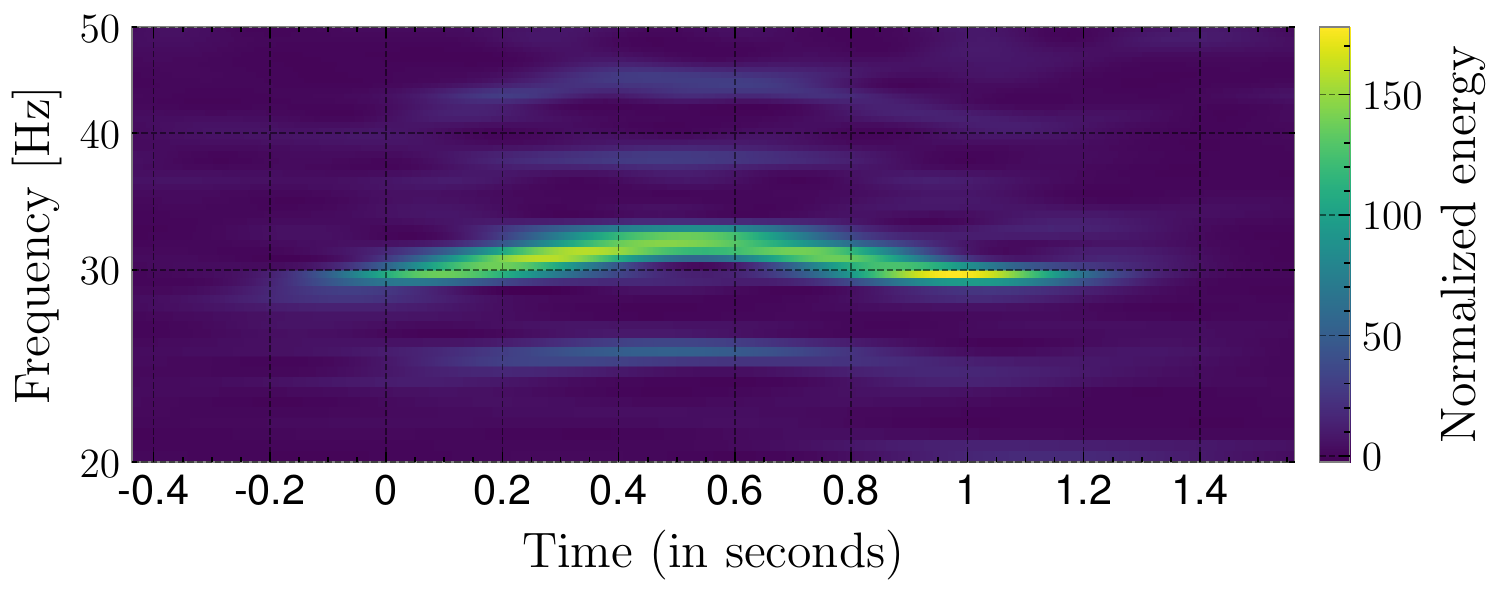}
         \caption{Scattered Light}
         \label{fig:scatteredlight}
     \end{subfigure}
     \hfill
     \begin{subfigure}[b]{0.4\textwidth}
         \centering
         \includegraphics[width=\columnwidth]{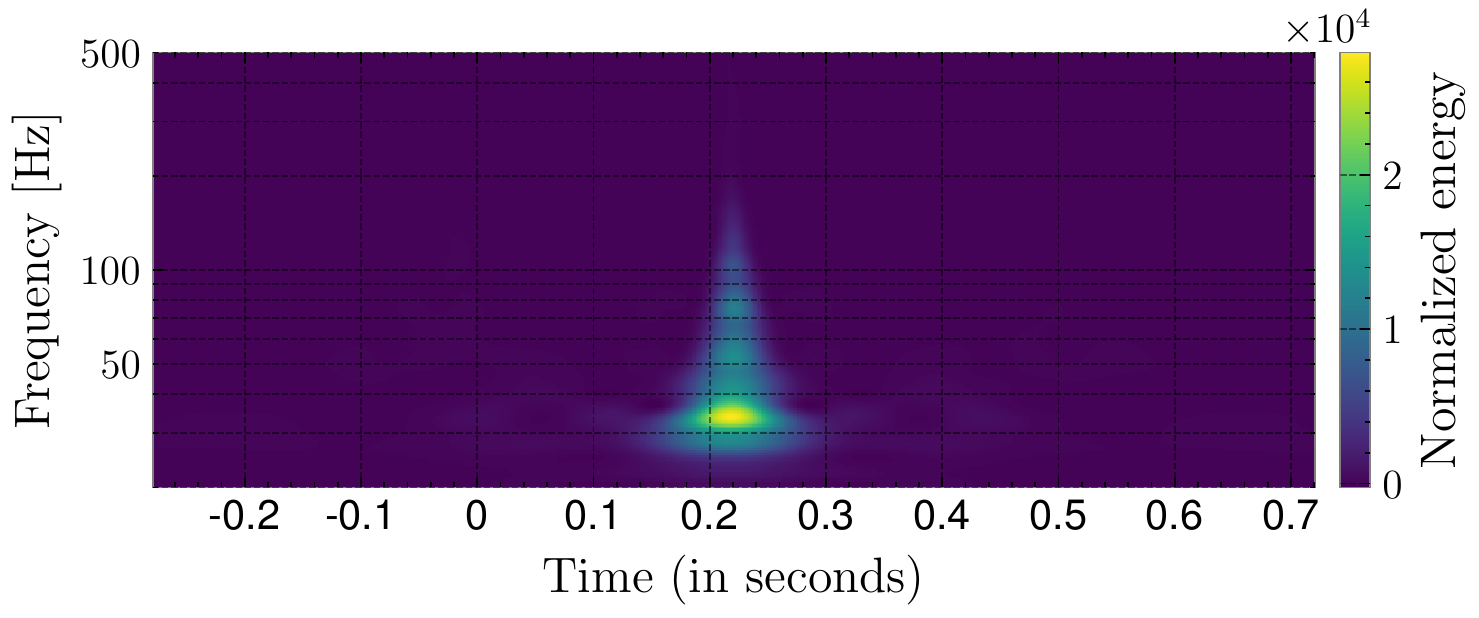}
         \caption{Koi Fish}
         \label{fig:koifish}
     \end{subfigure}
        \caption{Qscans of a few glitches commonly seen in the LVK data and a real gravitational wave signal GW190521}
        \label{fig:qscan}
\end{figure}

Fig \ref{fig:qscan} shows example qscans of the glitch classes we use in this work, along with the qscan of GW190521 \cite{LIGOScientific:2020iuh}, which due to its short temporal duration and loud nature, can be mis-identified as a tomte or low-frequency blip glitch \cite{Davis:2020nyf,jarov2023}. 

As the detectors become even more sensitive, this can lead to new glitches appearing in the data due to instrument modifications or glitches that were not loud enough to be seen with a higher ambient noise background. In addition to cases where the glitches have a low SNR, do not have auxiliary witnesses, or have very similar morphology to GW signals, it is useful to look for different ways to distinguish between signals and glitches that are robust to new glitch classes and manifestations.

This manuscript is organized as follows: In section \ref{Section 2} we review the parameter estimation methods used in this work to model glitch and simulated GW populations; section \ref{features} describes the set of features extracted from the posterior distributions; section \ref{rf} introduces the random forest classifier which is trained on this feature set to distinguish between glitches and GWs.

\section{Methodology} \label{Section 2}
We use the framework of Bayesian inference \cite{2019PASA...36...10T,Christensen:2022bxb} and random forests \cite{Breiman:2001hzm} to develop a classifier between signals and glitches. We use Bayesian inference to perform parameter estimation on both signals and glitches, while we use random forests for classification, based on a set of features (mentioned in section \ref{features}) that are extracted from the posteriors obtained through parameter estimation. We use the Bilby package \cite{2019ApJS..241...27A,2020MNRAS.499.3295R} to perform parameter estimation based on a GW signal model. The equation for the posterior probability distribution is given by

\begin{equation}
    p(\theta|\textbf{d}_i,h) = \frac{\mathcal{L}(\textbf{d}_i,h|\theta)\pi(\theta)}{\mathcal{Z}(\textbf{d}_i|h)},
\end{equation}
where $\theta$ are the model parameters and $h$ is the model, $p$ is called the posterior, $\mathcal{L}$ represents the likelihood, $\pi$ is the prior and $\mathcal{Z}$ is the marginalized likelihood, also called the evidence. The likelihood evaluation is based on the assumption that the noise is stationary and Gaussian \cite{whittle1951hypothesis}. Therefore, we use a stationary Gaussian noise model weighted by the detector power spectral density (PSD), expressed in frequency domain as \begin{equation}
    \textrm{ln} \mathcal{L}(\textbf{d}_i|\theta,h) \propto - \sum_{k}\frac{2|\tilde{d}_k - \tilde{h}_k(\theta)|^2}{S_n(f_k)T},
\end{equation}
where $f_k$ are the frequency bins, $d_k$ and $\tilde{h_k}$ are respectively the discrete Fourier transforms of the strain data
and waveform model, $S_n(f )$ is the noise PSD, and $T$ is
the duration of the analyzed data segment.
\subsection{Parameter Estimation}
The GW signature of a compact binary coalescence is completely defined by 15 parameters \cite{Veitch:2014wba}. There are 8 intrinsic parameters, out of which 2 are for specifying  the chirp mass $\mathcal{M}$ and mass ratio $q = \frac{m_2}{m_1}$ ($m_2<m_1$), and 6 for specifying the spin vectors; primary and secondary spin magnitudes $a_1,a_2$, and $t_1,t_2,\phi_{JL},\phi_{12}$ for specifying the orientation of the spin vectors. The other 7 are extrinsic parameters describing the luminosity distance to the binary $D_L$ and its orientation $\theta_{JN}$, polarization angle $\psi$, sky location in right ascension (ra, alpha) and declination (dec, delta) and the time and phase of coalescence $t_c,\phi_c$. If one or both of the components is a neutron star, there can be another 2 tidal deformation parameters $\Lambda_1,\Lambda_2$ as well. In this work, we use the IMRPhenomXPHM waveform family \cite{Pratten:2020ceb} as our waveform model. It is a fully-precessing inspiral-merger-ringdown model that includes higher-order multipole modes as well. To obtain the posterior distribution, we also need to specify a prior. To keep our analysis as unbiased as possible, we use uniform priors or astrophysically agnostic priors depending on the parameter. The priors are similar to ones used in Ashton et al.~\cite{Ashton:2021tvz}, including standard priors used in gravitational wave astronomy, as defined in Veitch et al.~\cite{Veitch:2014wba}. The non standard priors that we use are for chirp mass, for which we use a broad uniform prior in $[8,200]~\Msol$ to encompass all our simulated gravitational waves under one common prior. For the luminosity distance, we use the standard uniform in source frame prior between $[20,7500]\rm Mpc$. We used the same set of priors for signals and glitches for a fair comparison of their posteriors. Now that we have a model for our data and a prior, we can analyze it using a stochastic sampling algorithm. In this work we use the Bilby \cite{2019ApJS..241...27A,2020MNRAS.499.3295R} implementation of Dynesty \cite{2020MNRAS.493.3132S}, which is a nested sampling algorithm that provides both the posterior samples as well as the evidence.

\subsection{Glitches}
In this work, we focus on 5 classes of glitches that are frequently seen in the data and are known to decrease the search sensitivity \cite{Davis_2021,Davis:2020nyf,Cabero:2019orq,2021CQGra..38b5016S}. These glitches have been observed in all the previous Advanced LIGO observing runs and are likely to remain in the future as their sources have not yet been eliminated. Table \ref{table:1} lists the names of the glitch classes and the number of glitches in each class that we analyzed in this work. They are the same sample of glitches that were used in Ashton et al. \cite{Ashton:2021tvz}. These glitches were initially classified by the GravitySpy pipeline at a 95\% confidence threshold. We refer the reader to Ashton et al. \cite{Ashton:2021tvz} for more details on how the glitches were obtained from the Livingston \& Hanford detectors. Ashton et. al. \cite{Ashton:2021tvz} used the IMRPhenomPv2 waveform family \cite{Schmidt:2012rh,Hannam:2013oca} as the model for parameter estimation of their glitches. However, it does not include the effects of higher order modes which are known to have significant effects on posterior distributions, especially for systems with asymmetric masses. Since glitch posteriors tend to prefer high spins and $q<<1$ ratios \cite{Ashton:2021tvz}, we used the IMRPhenomXPHM waveform approximant which includes the effects of higher modes in this work.

\begin{table}
\centering
\begin{tabular}{||c|c|c||} 
 \hline
 Class & Hanford & Livingston \\ [0.5ex] 
 \hline\hline
 Blip & 987 & 962 \\ 
 \hline
 Tomte & 1000 & 1000 \\
 \hline
 Koi Fish & 92 & 93 \\
 \hline
 Fast Scattering & 100 & 100 \\
 \hline
 Scattered Light & 100 & 99 \\ [1ex] 
 \hline
\end{tabular}
\caption{The number of glitches in each glitch class from each LIGO detector that are used in this work}
\label{table:1}
\end{table}

\subsection{Gravitational Waves}
We simulated a large population of more than 300 gravitational waves each at LIGO Hanford and LIGO Livingston to create a complementary set to the glitches. We chose widely varying properties to cover the range of GW parameters that may be detectable by LIGO-Virgo-KAGRA in future observing runs. We used a single interferometer while performing parameter estimation as glitches would mostly appear in only one of the detectors. Hence it would be useful to compare glitches with single detector GW detection's. We used O3 sensitivity curves to simulate the noise in the detectors.  We chose primary and secondary masses from a uniformly spaced grid between $5~\Msol$ to $125~\Msol$. For each such combination, we assigned spin magnitudes $a_1$ and $a_2$ uniformly between 0 to 1. The rest of the parameters are assigned by sampling from their appropriate standard prior distributions \cite{Veitch:2014wba}. In particular, we sampled the luminosity distance from the uniform in comoving volume prior. Using these, we can compute the SNR for each binary. In case the SNR does not cross our detectable threshold of 8, we repeatedly sample from the luminosity distance prior and take the first value that does. Hence, all our simulated events lie inside the current detectable threshold for the detectors. We require this since we wish to build a diverse sample set of simulated GWs to complement the glitch set. It should be noted that this injection distribution is not astrophysically motivated \cite{KAGRA:2021duu}, but we follow this approach since our aim is to only create a classifier that can find statistical differences between glitch posteriors and those expected from gravitational waves.  Once we have a full set of simulated GW event parameters, we generate their corresponding waveforms using the IMRPhenomXPHM family. We perform parameter estimation on these injections using a similar set up as the glitches, except that the latter was done using data from the detectors while here we use synthetic noise. In addition to using the same priors for analysis, we also used the same time duration of 4 seconds of data for the parameter estimation of both glitches and simulated events for a fair comparison.

\begin{figure}
    \centering
    \includegraphics[width=\columnwidth]{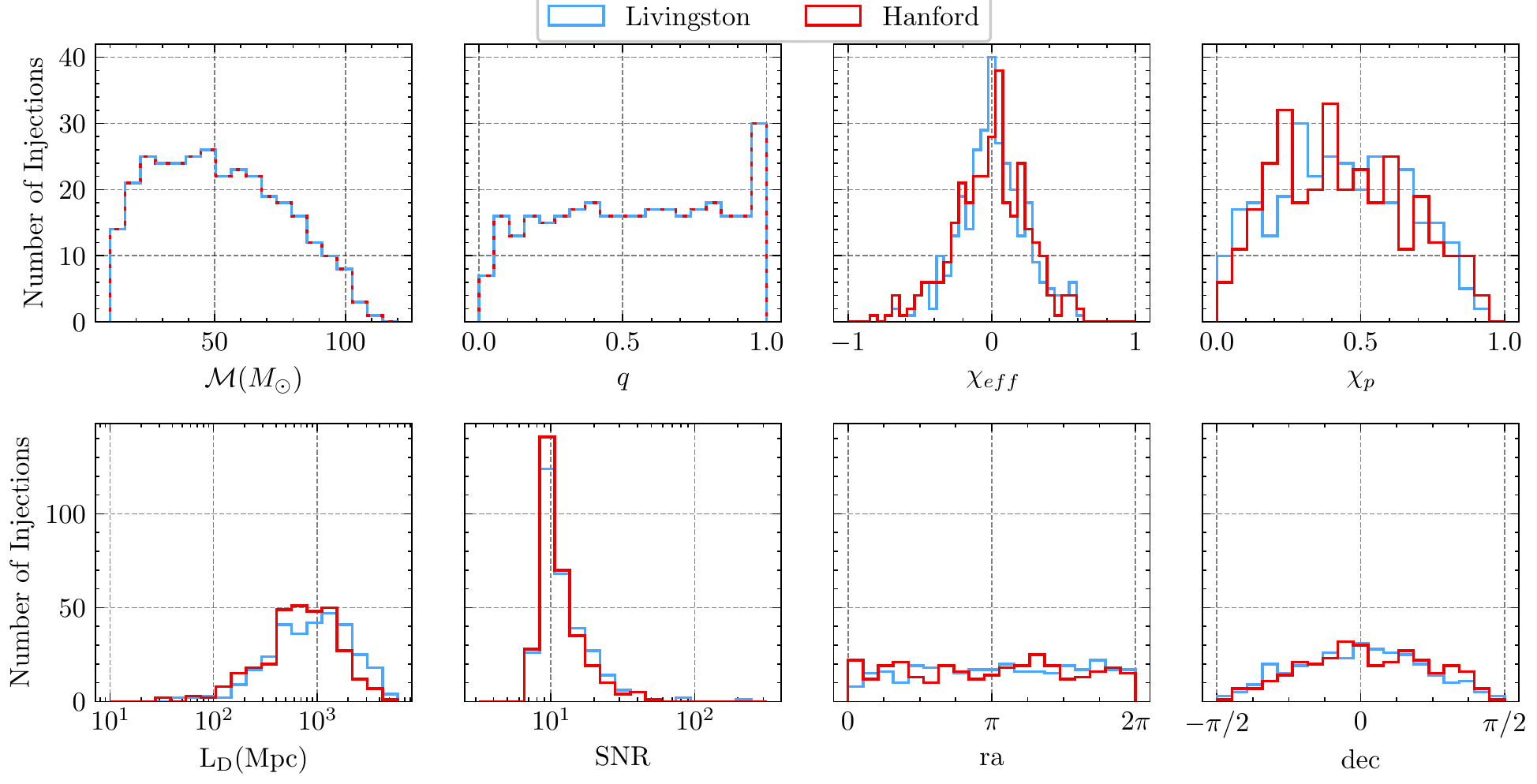}
    \caption{The distribution of parameters for the simulated GW signal population.}
    \label{fig:sim-distribution}
\end{figure}

\section{Features} \label{features}
It is computationally expensive to build a classifier using all the multi-dimensional posterior samples from each glitch and simulated GW. To reduce the complexity of our problem, we extract certain important features from the posterior distributions for different parameters and build a classifier using them. To incorporate the most likely values of a posterior distribution, we use the mode and median of the posterior samples for each parameter. To quantify the width in a distribution, we include both the $1\sigma$ (68 percentile) \& $2\sigma$ (95 percentile) intervals as features. We used the skew of the posterior distribution to quantify any asymmetries in them. 
Posterior distributions of different parameters may be correlated and these correlations may turn out to be different for glitches and gravitational waves. To include these effects, we also include pairwise linear correlation coefficients for each pair of parameters, which have a value of $+1$ for perfect correlation, $-1$ for anti-correlation, and $0$ for no correlation.

\section{Random forests} \label{rf}
The random forest \cite{Breiman:2001hzm} is an ensemble classifier that combines the result of many independent decision trees. The final classification is based on a majority voting of the decisions of all the constituent trees. The random forest was introduced partly to solve the overfitting problem to which individual decision trees are prone \cite{Breiman:2001hzm}. It has been extensively used for various classification problems in astronomy, including, but not limited to, variable star classification \cite{Richards}, photometric classification of supernovae \cite{staccato} and quasars \cite{RFquasars}, and sunspot classification \cite{sunspots}; it has also been used to improve gravitational wave data analysis \cite{Baker:2014eba}, to improve the LIGO detector duty cycle \cite{Biswas:2019wmx}, and to distinguish between NS-BH signals and noise transients in single-detector data \cite{Kapadia:2017fhb}. We choose the random forest as our classifier due to its ease of implementation, in addition to its robustness to noise and ability to assess input feature importance \cite{Breiman:2001hzm}.

The training samples for the random forest consists of the features extracted from the posterior distributions,  as described in Section \ref{features}. For each glitch in a glitch class, or a simulated gravitational wave, we have an n-tuple of numbers corresponding to the values of the posterior features for all parameters. It should be noted that our training data sets have some class imbalance, as we have more total glitches than simulated gravitational waves and even a difference in the number of glitches in each glitch class. To account for this, we use SMOTE (Synthetic minority oversampling technique) \cite{Chawla_2002} which balances the training data set by generating extra samples in a way similar to bootstrap without adding any new information or bias to the training set.

It should be noted that our sample data has around 200 features for each posterior distribution including mean, median, widths, skew for all parameters and their pairwise linear correlation coefficients. A benefit to using a random forest is that we do not need to perform feature selection, as this is done by the algorithm when a random subset of features is obtained to be used for splitting at each node of each tree in the ensemble; such ``bagging`` of the features in addition to the trees themselves helps prevent overfitting \cite{Breiman:2001hzm}.


We set up the random forest based on a few commonly used hyper-parameters which are the number of trees, the criterion for creating a split, and the maximum depth. We use $100$ trees and the gini entropy criterion to measure the quality of a split. The trees are extended till all the leaf nodes are pure, i.e they consist of samples of only a single class. We use a $70-30$ train-test split on our sample data to create and test the classifier.

%


\begin{figure}
     \centering
     \begin{subfigure}[b]{\textwidth}
         \centering
         \includegraphics[width=\columnwidth]{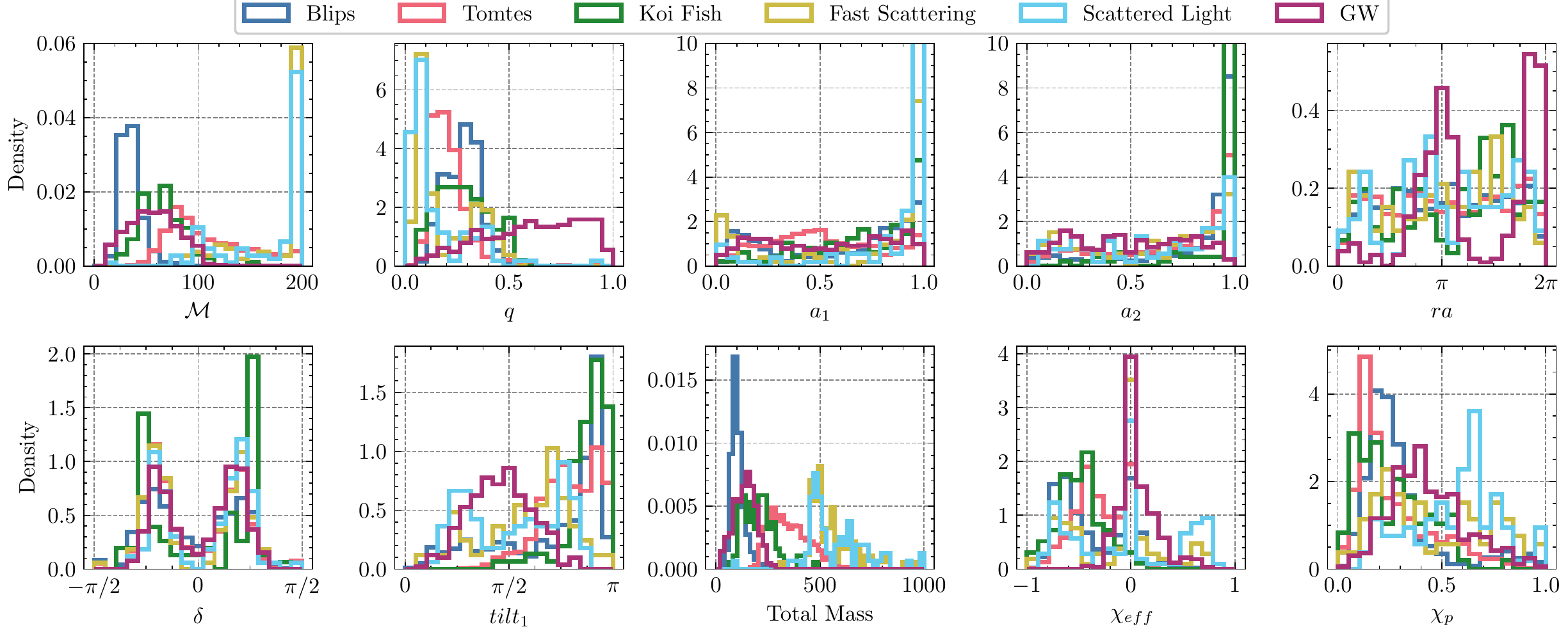}
         \caption{Distribution of posterior modes of glitches and simulated GWs at Hanford}
         \label{fig:glitch_modes_LHO}
     \end{subfigure}
     \hfill
     \begin{subfigure}[b]{\textwidth}
         \centering
         \includegraphics[width=\columnwidth]{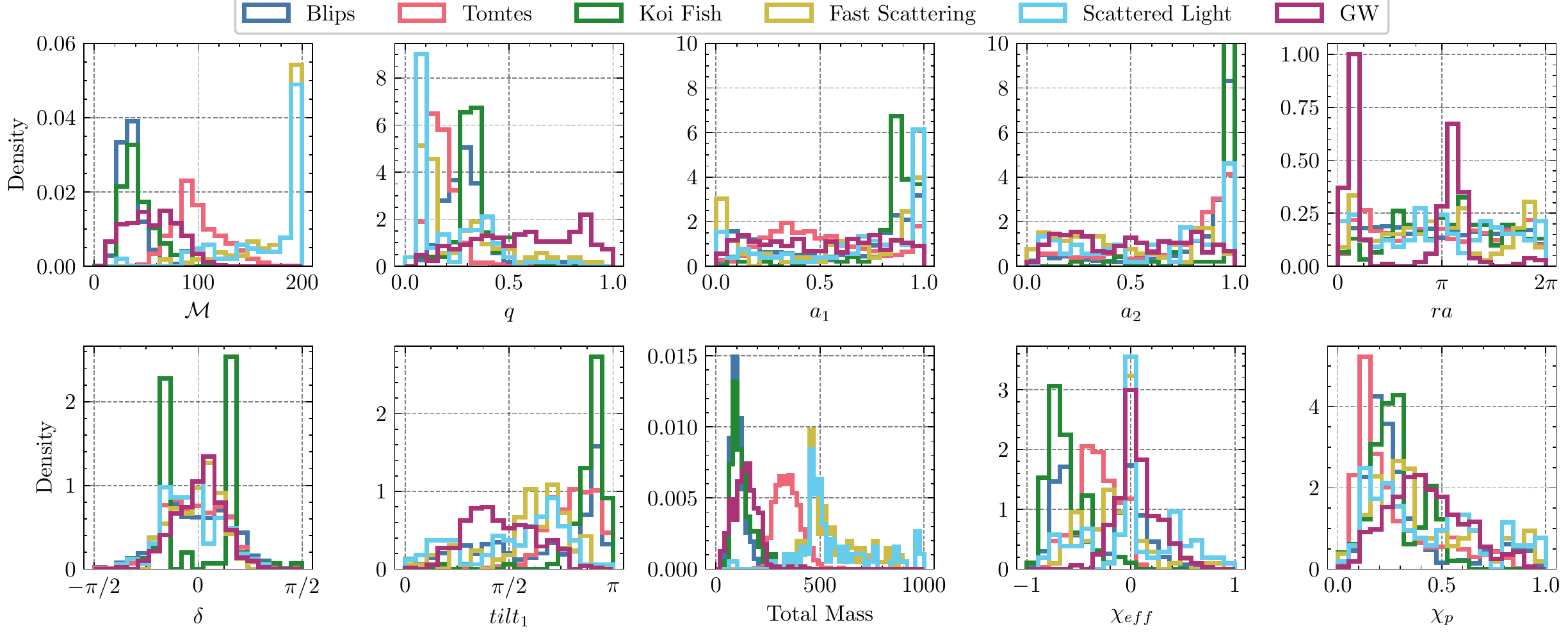}
         \caption{Distribution of posterior modes of glitches and simulated GWs at Livingston}
         \label{fig:glitch_modes_LLO}
     \end{subfigure}
    \caption{The above figures show the histogram of the modes that are used as features for the random forest classifier for some of the parameters}
    \label{modes_histogram}
\end{figure}

\section{Results}
We divide this section into 2 parts. The first one describes the trends seen in the posterior distributions for glitches of different classes, and how they might differ from what the posteriors of real signals would look like. In the second part, we show the results of our random forest classifier, trained on the features extracted from the posteriors of glitches and simulated signals.

\subsection{Glitch vs signal posterior distributions}
In Fig \ref{modes_histogram}, we notice that for each glitch class, the posteriors for the individual glitches tend to cluster around a narrow range of values for a few parameters, which is consistent with the results in Ashton et. al. It is most commonly seen in their chirp mass, mass ratio, total mass, spin magnitude posteriors as well as a few other parameters that vary depending on the glitch class. We also notice that the widths of the posteriors tend to be narrower for glitches than for simulated gravitational waves. Another feature, that was also seen in Ashton et al. \cite{Ashton:2021tvz}, is that glitches tend to prefer high spin posteriors and asymmetric mass ratios ($q \ll 1$). This differs quite significantly from the posteriors of simulated gravitational waves, which often tend to be uninformative in spin magnitude. However, it should be noted that posterior widths can be correlated with how strong the signal is, hence very high SNR gravitational wave events would have narrow posteriors as well. We also see a similar behaviour in various other parameters like $\chi_{\rm eff},\chi_{\rm p},\textrm{tilt}$ angles, where the posteriors of simulated events are uninformative (i.e they reproduce the prior), while glitches tend to prefer certain specific regions for those parameters.

\subsection{Classification}
\begin{figure}
     \centering
     \begin{subfigure}[b]{\textwidth}
         \centering
         \includegraphics[width=\columnwidth]{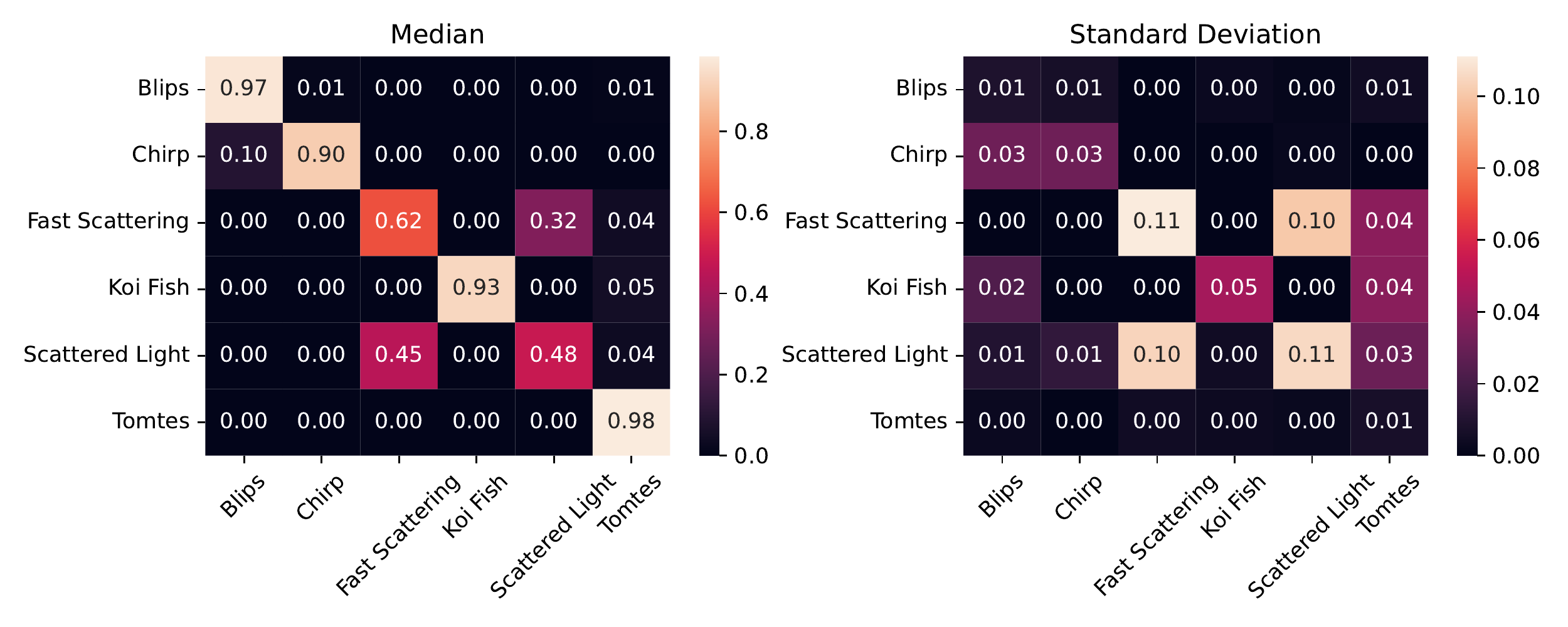}
         \caption{Median confusion matrix of the random forest classifier trained and tested using all features of the Hanford set along with the standard deviations for the corresponding confusion matrix values}
         \label{fig:all_features_h}
     \end{subfigure}
     \hfill
     \begin{subfigure}[b]{\textwidth}
         \centering
         \includegraphics[width=\columnwidth]{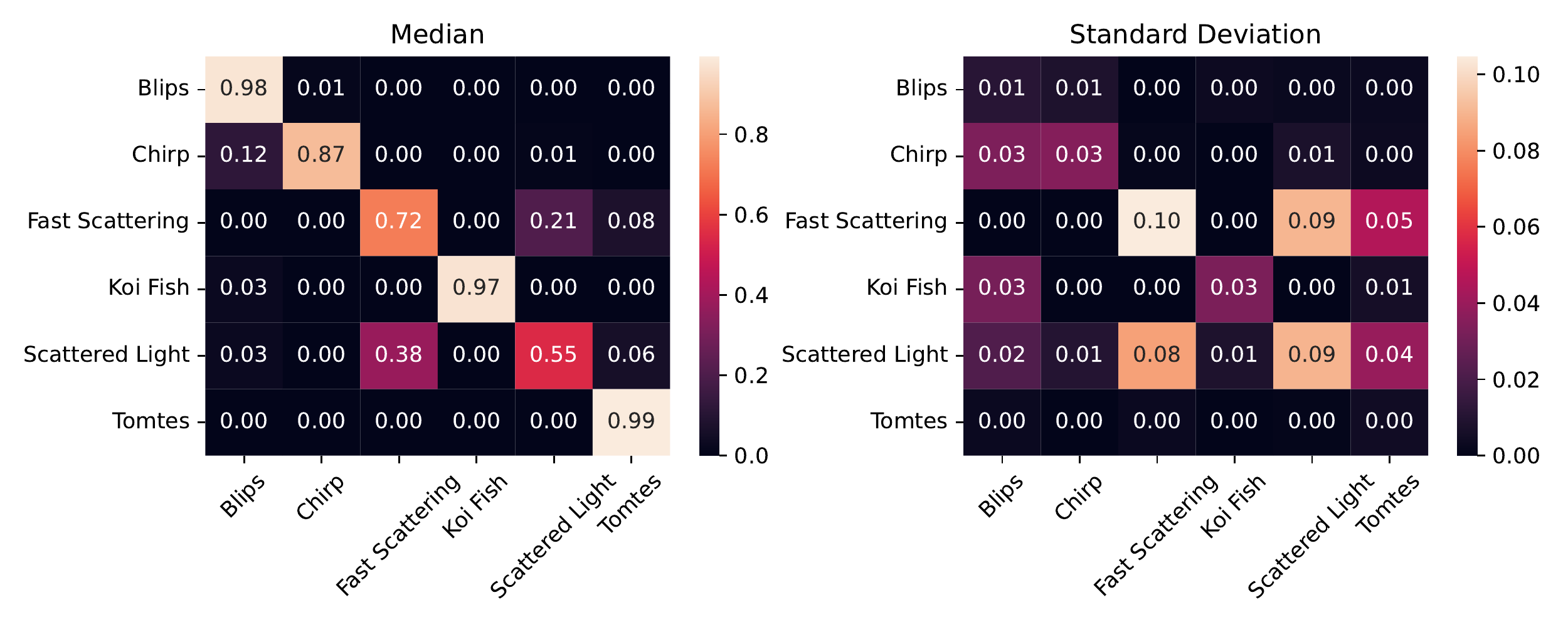}
         \caption{Median confusion matrix of the random forest classifier trained and tested using all features of the Livingston set along with the standard deviations for the corresponding confusion matrix values}
         \label{fig:all_features_l}
     \end{subfigure}
     \caption{Results of the random forest classifier trained using all features}
     \label{all_features}
\end{figure}

\begin{figure}
     \centering
     \begin{subfigure}[b]{\textwidth}
         \centering
         \includegraphics[width=\columnwidth]{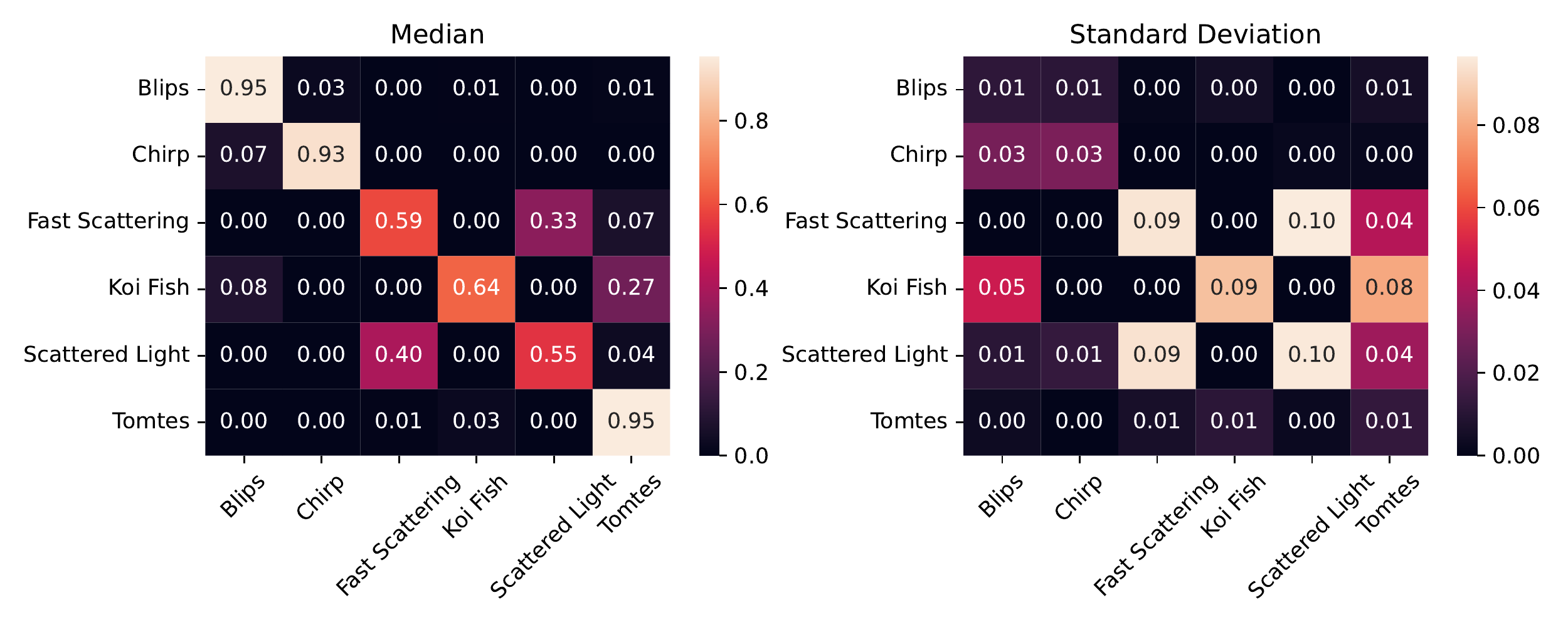}
         \caption{Median confusion matrix of the random forest classifier trained and tested using only modes of the Hanford set along with the standard deviations for the corresponding confusion matrix values}
         \label{fig:all_features_h_mode}
     \end{subfigure}
     \hfill
     \begin{subfigure}[b]{\textwidth}
         \centering
         \includegraphics[width=\columnwidth]{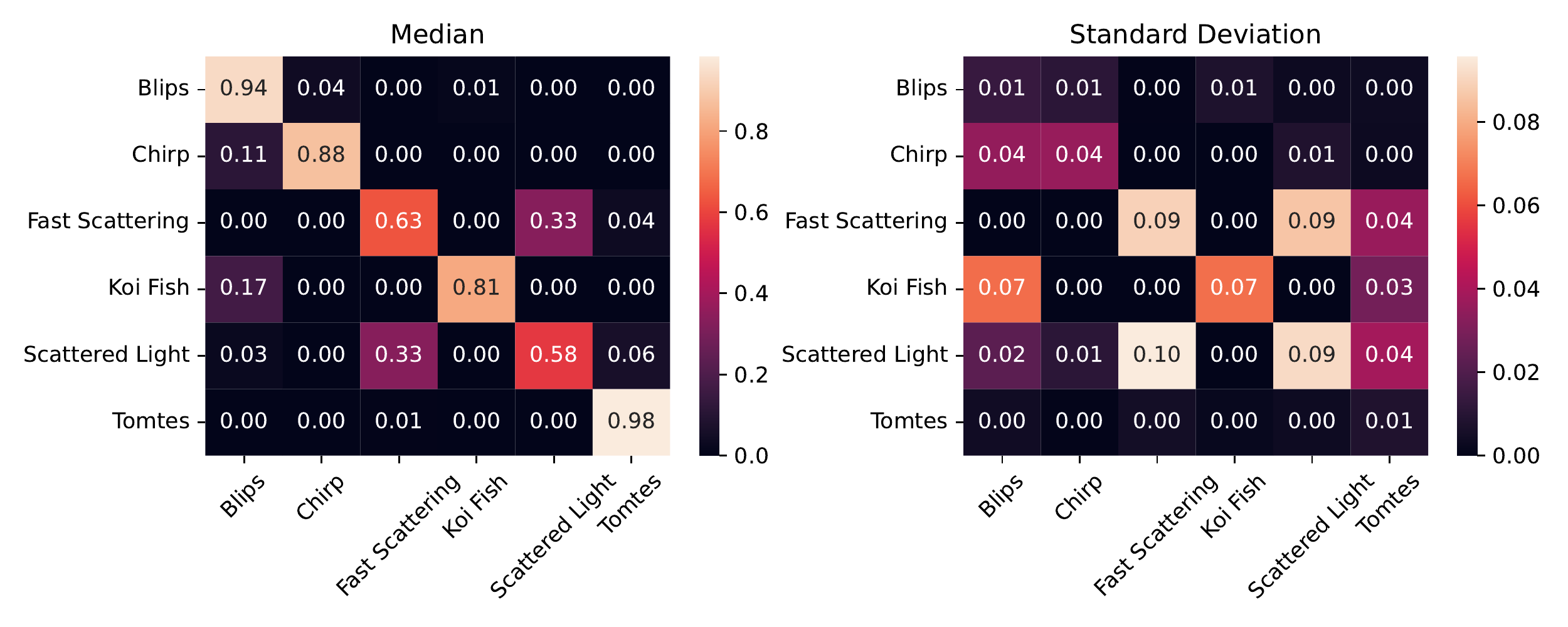}
         \caption{Median confusion matrix of the random forest classifier trained and tested using only modes of the Livingston set along with the standard deviations for the corresponding confusion matrix values}
         \label{fig:all_features_l_mode}
     \end{subfigure}
     \caption{Results of the random forest classifier trained using only modes}
     \label{modes}
\end{figure}

We train a random forest classifier for the glitches and simulated signals using the Hanford and Livingston sets separately. Due to the inherent randomness involved in training and testing random forests, we repeat this process 100 times for each of the detectors and report the median values in the confusion matrix along with the corresponding standard deviations. This would give a more accurate picture of the performance of the classifier than a single iteration of it. Since there is a class imbalance between the classes, we used SMOTE \cite{Chawla_2002} to balance them in the training set. The results look promising, as seen in Fig \ref{all_features}. We obtain a high mean accuracy averaged over all classes of 0.93 for the Hanford classifier and 0.94 for the Livingston classifier with a standard deviation of $<0.01$ across all the iterations. We also achieve a mean chirp recall of $0.90$ for the Hanford classifier and $0.88$ for the Livingston classifier with standard deviations of $0.03$ correspondingly in the chirp recall values across the 100 iterations. This means that not only are most of the glitches classified correctly and not misidentified as a gravitational wave, but we also have very few false negatives in identifying gravitational waves, which is essential as one of our top priorities is not rejecting true GW signals. We also notice that the classifier has a hard time in distinguishing between fast scattering and scattered light glitches. This is expected as their posteriors show quite similar distributions, hence their extracted features would be similar. This is not a problem as our aim is to build a GW vs Glitch classifier and not classification within different glitch families. We also find out that the most essential features that were useful for the random forest to make classifications were the modes and medians of chirp mass and total mass, medians of mass ratio and $2\sigma$ posterior widths of total mass, the $\textrm{tilt}\_2$ angle and $\mathcal{M}$. In general, the credible intervals of the marginalized posteriors for simulated gravitational waves tend to be broader than those of glitches, as all the glitches in our set have a very high SNR. The regions where the classifier might fail, that is, predict a real gravitational wave as a type of glitch is when the signal would be such that its posteriors lie in the regions where the posteriors of a type of glitch class tend to cluster. However, since we not only take into account the medians of the posterior, merely lying in the same region may not be a problem as the random forest would also use other features such as posterior widths for classification, and those would most likely not be similar. 
We additionally train another random forest classifier but just using modes as the feature set this time around. The resulting median confusion matrix is shown in Fig \ref{modes}. We obtain quite similar results as when it was trained on all features, which shows that the modes are the most essential features to classify between different types of glitches and gravitational waves. This is a particularly useful result, as it hints towards a similar type of random forest classifier that can be created using the matched filtering estimates of the astrophysical parameters of a candidate signal, which can be computed much faster that the posterior samples in parameter estimation. If there indeed turns out to be clusterings in the matched filter estimates for different types of glitches, a similar type of random forest classifier can be created and used in the low-latency pipeline for quick classifications. However, only a detailed study can verify if this would be possible.

\section{Conclusion}
In this work, we created a simulated set of gravitational waves and a set of glitches belonging to 5 glitch classes that were obtained from Ashton et al \cite{Ashton:2021tvz}. We used Bayesian inference to perform parameter estimation on them using a precessing waveform model, that also includes the effects of higher modes. We extracted certain broad features from each parameter estimation run to efficiently describe the population properties of the posteriors of glitches and gravitational waves. We trained a random forest using the extracted features, and found that it has a high test accuracy as well a low false negative rate for classifying GWs. This could be a useful technique to validate single detector candidate detection's \cite{Callister:2017urp,Nitz:2020naa,Davies:2022thw}, either due to the other detectors being offline, or having a low SNR due to their orientation or sensitivity. This method could be used to validate candidate signals offline to assess which glitch class or gravitational wave they might belong to. This method could also find an application in online matched filter searches, where a random forest classifier is trained on the matched filtering parameter estimates of candidate detection's. This could be used in addition to existing methods to validate signals, thereby possibly increasing the confidence of detected signals or improving the reach of the detectors. 

\section{Acknowledgments}
This material is based upon work supported by NSF's LIGO Laboratory which is a major facility fully funded by the National Science Foundation. Virgo is funded by the French Centre National de  Recherche  Scientifique  (CNRS),  the  Italian  Istituto Nazionale  della  Fisica  Nucleare  (INFN)  and  the  Dutch Nikhef, with contributions by Polish and Hungarian institutes.   We  are  also  grateful  to  computing  resources provided  by  the  LIGO  Laboratory  computing  clusters at California Institute of Technology and LIGO Hanford Observatory supported by National Science Foundation Grants PHY-0757058 and PHY-0823459.  The majority of  analysis  performed  for  this  research  was  done  using resources  provided  by  the  Open  Science  Grid (https://osg-htc.org/) \cite{Pordes:2007zzb,5171374}, which is supported by the National Science Foundation award \#2030508. This research was enabled in part by support provided by the BC DRI Group and the Digital Research Alliance of Canada (alliancecan.ca).
This work makes use of the SciPy \cite{2020SciPy-NMeth}, NumPy \cite{harris2020array}, GWpy \cite{gwpy}, and PyCBC \cite{Nitz:2017svb} software for data analysis and visualisation. 
NS is supported by the
Mitacs Globalink Research Internship, JM by the Canada Research Chairs program, and AMK by a Killam Doctoral Scholarship. DCS. is supported by National Science and Engineering Research Council of Canada grant number RGPIN/03985-2021.

\printbibliography

@ARTICLE{2019PASA...36...10T,
       author = {{Thrane}, Eric and {Talbot}, Colm},
        title = "{An introduction to Bayesian inference in gravitational-wave astronomy: Parameter estimation, model selection, and hierarchical models}",
      journal = {\pasa},
         year = 2019,
        month = mar,
       volume = {36},
          eid = {e010},
        pages = {e010},
          doi = {10.1017/pasa.2019.2},
archivePrefix = {arXiv},
       eprint = {1809.02293},
 primaryClass = {astro-ph.IM},
       adsurl = {https://ui.adsabs.harvard.edu/abs/2019PASA...36...10T}
}

@article{Pratten:2020ceb,
    author = "Pratten, Geraint and others",
    title = "{Computationally efficient models for the dominant and subdominant harmonic modes of precessing binary black holes}",
    eprint = "2004.06503",
    archivePrefix = "arXiv",
    primaryClass = "gr-qc",
    doi = "10.1103/PhysRevD.103.104056",
    journal = "Phys. Rev. D",
    volume = "103",
    number = "10",
    pages = "104056",
    year = "2021"
}

@article{LIGO:2021ppb,
    author = "Davis, Derek and others",
    collaboration = "LIGO",
    title = "{LIGO detector characterization in the second and third observing runs}",
    eprint = "2101.11673",
    archivePrefix = "arXiv",
    primaryClass = "astro-ph.IM",
    reportNumber = "P2000495",
    doi = "10.1088/1361-6382/abfd85",
    journal = "Class. Quant. Grav.",
    volume = "38",
    number = "13",
    pages = "135014",
    year = "2021"
}

@article{Ashton:2021tvz,
    author = "Ashton, Gregory and Thiele, Sarah and Lecoeuche, Yannick and McIver, Jess and Nuttall, Laura K.",
    title = "{Parameterised population models of transient non-Gaussian noise in the LIGO gravitational-wave detectors}",
    eprint = "2110.02689",
    archivePrefix = "arXiv",
    primaryClass = "gr-qc",
    doi = "10.1088/1361-6382/ac8094",
    journal = "Class. Quant. Grav.",
    volume = "39",
    number = "17",
    pages = "175004",
    year = "2022"
}

@article{Hannam:2013oca,
    author = {Hannam, Mark and Schmidt, Patricia and Boh\'e, Alejandro and Haegel, Le\"\i{}la and Husa, Sascha and Ohme, Frank and Pratten, Geraint and P\"urrer, Michael},
    title = "{Simple Model of Complete Precessing Black-Hole-Binary Gravitational Waveforms}",
    eprint = "1308.3271",
    archivePrefix = "arXiv",
    primaryClass = "gr-qc",
    doi = "10.1103/PhysRevLett.113.151101",
    journal = "Phys. Rev. Lett.",
    volume = "113",
    number = "15",
    pages = "151101",
    year = "2014"
}

@article{Schmidt:2012rh,
    author = "Schmidt, Patricia and Hannam, Mark and Husa, Sascha",
    title = "{Towards models of gravitational waveforms from generic binaries: A simple approximate mapping between precessing and non-precessing inspiral signals}",
    eprint = "1207.3088",
    archivePrefix = "arXiv",
    primaryClass = "gr-qc",
    doi = "10.1103/PhysRevD.86.104063",
    journal = "Phys. Rev. D",
    volume = "86",
    pages = "104063",
    year = "2012"
}

@ARTICLE{2020MNRAS.493.3132S,
       author = {{Speagle}, Joshua S.},
        title = "{DYNESTY: a dynamic nested sampling package for estimating Bayesian posteriors and evidences}",
      journal = {\mnras},
         year = 2020,
        month = apr,
       volume = {493},
       number = {3},
        pages = {3132-3158},
          doi = {10.1093/mnras/staa278},
archivePrefix = {arXiv},
       eprint = {1904.02180},
 primaryClass = {astro-ph.IM},
       adsurl = {https://ui.adsabs.harvard.edu/abs/2020MNRAS.493.3132S}
}

@article{Veitch:2014wba,
    author = "Veitch, J. and others",
    title = "{Parameter estimation for compact binaries with ground-based gravitational-wave observations using the LALInference software library}",
    eprint = "1409.7215",
    archivePrefix = "arXiv",
    primaryClass = "gr-qc",
    reportNumber = "LIGO-P1400152",
    doi = "10.1103/PhysRevD.91.042003",
    journal = "Phys. Rev. D",
    volume = "91",
    number = "4",
    pages = "042003",
    year = "2015"
}

@article{Breiman:2001hzm,
    author = "Breiman, Leo",
    title = "{Random Forests}",
    doi = "10.1023/A:1010933404324",
    journal = "Machine Learning",
    volume = "45",
    number = "1",
    pages = "5--32",
    year = "2001"
}

@ARTICLE{2019ApJS..241...27A,
       author = {{Ashton}, Gregory and {H{\"u}bner}, Moritz and {Lasky}, Paul D. and {Talbot}, Colm and {Ackley}, Kendall and {Biscoveanu}, Sylvia and {Chu}, Qi and {Divakarla}, Atul and {Easter}, Paul J. and {Goncharov}, Boris and {Hernandez Vivanco}, Francisco and {Harms}, Jan and {Lower}, Marcus E. and {Meadors}, Grant D. and {Melchor}, Denyz and {Payne}, Ethan and {Pitkin}, Matthew D. and {Powell}, Jade and {Sarin}, Nikhil and {Smith}, Rory J.~E. and {Thrane}, Eric},
        title = "{BILBY: A User-friendly Bayesian Inference Library for Gravitational-wave Astronomy}",
      journal = {\apjs},
         year = 2019,
        month = apr,
       volume = {241},
       number = {2},
          eid = {27},
        pages = {27},
          doi = {10.3847/1538-4365/ab06fc},
archivePrefix = {arXiv},
       eprint = {1811.02042},
 primaryClass = {astro-ph.IM},
       adsurl = {https://ui.adsabs.harvard.edu/abs/2019ApJS..241...27A}
}

@ARTICLE{2020MNRAS.499.3295R,
       author = {{Romero-Shaw}, I.~M. and {Talbot}, C. and {Biscoveanu}, S. and {D'Emilio}, V. and {Ashton}, G. and {Berry}, C.~P.~L. and {Coughlin}, S. and {Galaudage}, S. and {Hoy}, C. and {H{\"u}bner}, M. and {Phukon}, K.~S. and {Pitkin}, M. and {Rizzo}, M. and {Sarin}, N. and {Smith}, R. and {Stevenson}, S. and {Vajpeyi}, A. and {Ar{\`e}ne}, M. and {Athar}, K. and {Banagiri}, S. and {Bose}, N. and {Carney}, M. and {Chatziioannou}, K. and {Clark}, J.~A. and {Colleoni}, M. and {Cotesta}, R. and {Edelman}, B. and {Estell{\'e}s}, H. and {Garc{\'\i}a-Quir{\'o}s}, C. and {Ghosh}, Abhirup and {Green}, R. and {Haster}, C. -J. and {Husa}, S. and {Keitel}, D. and {Kim}, A.~X. and {Hernandez-Vivanco}, F. and {Maga{\~n}a Hernandez}, I. and {Karathanasis}, C. and {Lasky}, P.~D. and {De Lillo}, N. and {Lower}, M.~E. and {Macleod}, D. and {Mateu-Lucena}, M. and {Miller}, A. and {Millhouse}, M. and {Morisaki}, S. and {Oh}, S.~H. and {Ossokine}, S. and {Payne}, E. and {Powell}, J. and {Pratten}, G. and {P{\"u}rrer}, M. and {Ramos-Buades}, A. and {Raymond}, V. and {Thrane}, E. and {Veitch}, J. and {Williams}, D. and {Williams}, M.~J. and {Xiao}, L.},
        title = "{Bayesian inference for compact binary coalescences with BILBY: validation and application to the first LIGO-Virgo gravitational-wave transient catalogue}",
      journal = {\mnras},
         year = 2020,
        month = dec,
       volume = {499},
       number = {3},
        pages = {3295-3319},
          doi = {10.1093/mnras/staa2850},
archivePrefix = {arXiv},
       eprint = {2006.00714},
 primaryClass = {astro-ph.IM},
       adsurl = {https://ui.adsabs.harvard.edu/abs/2020MNRAS.499.3295R}
}

@ARTICLE{2017CQGra..34f4003Z,
       author = {{Zevin}, M. and {Coughlin}, S. and {Bahaadini}, S. and {Besler}, E. and {Rohani}, N. and {Allen}, S. and {Cabero}, M. and {Crowston}, K. and {Katsaggelos}, A.~K. and {Larson}, S.~L. and {Lee}, T.~K. and {Lintott}, C. and {Littenberg}, T.~B. and {Lundgren}, A. and {{\O}sterlund}, C. and {Smith}, J.~R. and {Trouille}, L. and {Kalogera}, V.},
        title = "{Gravity Spy: integrating advanced LIGO detector characterization, machine learning, and citizen science}",
      journal = {Classical and Quantum Gravity},
         year = 2017,
        month = mar,
       volume = {34},
       number = {6},
          eid = {064003},
        pages = {064003},
          doi = {10.1088/1361-6382/aa5cea},
archivePrefix = {arXiv},
       eprint = {1611.04596},
 primaryClass = {gr-qc},
       adsurl = {https://ui.adsabs.harvard.edu/abs/2017CQGra..34f4003Z}
}

@article{LIGOScientific:2016gtq,
    author = "Abbott, B. P. and others",
    collaboration = "LIGO Scientific, Virgo",
    title = "{Characterization of transient noise in Advanced LIGO relevant to gravitational wave signal GW150914}",
    eprint = "1602.03844",
    archivePrefix = "arXiv",
    primaryClass = "gr-qc",
    doi = "10.1088/0264-9381/33/13/134001",
    journal = "Class. Quant. Grav.",
    volume = "33",
    number = "13",
    pages = "134001",
    year = "2016"
}

@article{Nuttall:2015dqa,
    author = "Nuttall, L. and others",
    title = "{Improving the Data Quality of Advanced LIGO Based on Early Engineering Run Results}",
    eprint = "1508.07316",
    archivePrefix = "arXiv",
    primaryClass = "gr-qc",
    doi = "10.1088/0264-9381/32/24/245005",
    journal = "Class. Quant. Grav.",
    volume = "32",
    number = "24",
    pages = "245005",
    year = "2015"
}

@article{LIGOScientific:2017tza,
    author = "Abbott, B P and others",
    collaboration = "LIGO Scientific, Virgo",
    title = "{Effects of data quality vetoes on a search for compact binary coalescences in Advanced LIGO\textquoteright{}s first observing run}",
    eprint = "1710.02185",
    archivePrefix = "arXiv",
    primaryClass = "gr-qc",
    doi = "10.1088/1361-6382/aaaafa",
    journal = "Class. Quant. Grav.",
    volume = "35",
    number = "6",
    pages = "065010",
    year = "2018"
}

@article{LIGOScientific:2018mvr,
    author = "Abbott, B. P. and others",
    collaboration = "LIGO Scientific, Virgo",
    title = "{GWTC-1: A Gravitational-Wave Transient Catalog of Compact Binary Mergers Observed by LIGO and Virgo during the First and Second Observing Runs}",
    eprint = "1811.12907",
    archivePrefix = "arXiv",
    primaryClass = "astro-ph.HE",
    reportNumber = "LIGO-P1800307",
    doi = "10.1103/PhysRevX.9.031040",
    journal = "Phys. Rev. X",
    volume = "9",
    number = "3",
    pages = "031040",
    year = "2019"
}

@article{LIGOScientific:2020ibl,
    author = "Abbott, R. and others",
    collaboration = "LIGO Scientific, Virgo",
    title = "{GWTC-2: Compact Binary Coalescences Observed by LIGO and Virgo During the First Half of the Third Observing Run}",
    eprint = "2010.14527",
    archivePrefix = "arXiv",
    primaryClass = "gr-qc",
    reportNumber = "P2000061",
    doi = "10.1103/PhysRevX.11.021053",
    journal = "Phys. Rev. X",
    volume = "11",
    pages = "021053",
    year = "2021"
}

@article{LIGOScientific:2021djp,
    author = "Abbott, R. and others",
    collaboration = "LIGO Scientific, VIRGO, KAGRA",
    title = "{GWTC-3: Compact Binary Coalescences Observed by LIGO and Virgo During the Second Part of the Third Observing Run}",
    eprint = "2111.03606",
    archivePrefix = "arXiv",
    primaryClass = "gr-qc",
    reportNumber = "LIGO-P2000318",
    month = "11",
    year = "2021"
}

@article{Essick:2020qpo,
    author = "Essick, Reed and Godwin, Patrick and Hanna, Chad and Blackburn, Lindy and Katsavounidis, Erik",
    title = "{iDQ: Statistical Inference of Non-Gaussian Noise with Auxiliary Degrees of Freedom in Gravitational-Wave Detectors}",
    eprint = "2005.12761",
    archivePrefix = "arXiv",
    primaryClass = "astro-ph.IM",
    month = "5",
    year = "2020"
}

@article{Godwin:2020weu,
    author = "Godwin, Patrick and others",
    title = "{Incorporation of Statistical Data Quality Information into the GstLAL Search Analysis}",
    eprint = "2010.15282",
    archivePrefix = "arXiv",
    primaryClass = "gr-qc",
    month = "10",
    year = "2020"
}

@article{Chatterjee:2019avs,
    author = "Chatterjee, Deep and Ghosh, Shaon and Brady, Patrick R. and Kapadia, Shasvath J. and Miller, Andrew L. and Nissanke, Samaya and Pannarale, Francesco",
    title = "{A Machine Learning Based Source Property Inference for Compact Binary Mergers}",
    eprint = "1911.00116",
    archivePrefix = "arXiv",
    primaryClass = "astro-ph.IM",
    doi = "10.3847/1538-4357/ab8dbe",
    journal = "Astrophys. J.",
    volume = "896",
    number = "1",
    pages = "54",
    year = "2020"
}

@article{PhysRevD.71.062001,
  title = {${\ensuremath{\chi}}^{2}$ time-frequency discriminator for gravitational wave detection},
  author = {Allen, Bruce},
  journal = {Phys. Rev. D},
  volume = {71},
  issue = {6},
  pages = {062001},
  numpages = {22},
  year = {2005},
  month = {Mar},
  publisher = {American Physical Society},
  doi = {10.1103/PhysRevD.71.062001},
  url = {https://link.aps.org/doi/10.1103/PhysRevD.71.062001}
}

@article{Chatterji:2004qg,
    author = "Chatterji, S. and Blackburn, L. and Martin, G. and Katsavounidis, E.",
    title = "{Multiresolution techniques for the detection of gravitational-wave bursts}",
    eprint = "gr-qc/0412119",
    archivePrefix = "arXiv",
    doi = "10.1088/0264-9381/21/20/024",
    journal = "Class. Quant. Grav.",
    volume = "21",
    pages = "S1809--S1818",
    year = "2004"
}

@article{Davis:2020nyf,
    author = "Davis, Derek and White, Laurel V. and Saulson, Peter R.",
    title = "{Utilizing aLIGO Glitch Classifications to Validate Gravitational-Wave Candidates}",
    eprint = "2002.09429",
    archivePrefix = "arXiv",
    primaryClass = "gr-qc",
    doi = "10.1088/1361-6382/ab91e6",
    journal = "Class. Quant. Grav.",
    volume = "37",
    number = "14",
    pages = "145001",
    year = "2020"
}

@article{Virgo:2022fxr,
    author = "Acernese, F. and others",
    collaboration = "Virgo",
    title = "{Virgo Detector Characterization and Data Quality during the O3 run}",
    eprint = "2205.01555",
    archivePrefix = "arXiv",
    primaryClass = "gr-qc",
    month = "5",
    year = "2022"
}

@article{Chawla_2002,
	doi = {10.1613/jair.953},
  
	url = {https://doi.org/10.1613\%2Fjair.953},
  
	year = 2002,
	month = {jun},
  
	publisher = {{AI} Access Foundation},
  
	volume = {16},
  
	pages = {321--357},
  
	author = {N. V. Chawla and K. W. Bowyer and L. O. Hall and W. P. Kegelmeyer},
  
	title = {{SMOTE}: Synthetic Minority Over-sampling Technique},
  
	journal = {Journal of Artificial Intelligence Research}
}

@article{Davis_2021,
  title = {Data Quality Vetoes Applied to the Analysis of LIGO Data from the Second Observing Run},
  year = {2021},
  month = {May},
  url = {https://dcc.ligo.org/LIGO-T2100150}
}

@misc{sachdev2019gstlal,
      title={The GstLAL Search Analysis Methods for Compact Binary Mergers in Advanced LIGO's Second and Advanced Virgo's First Observing Runs}, 
      author={Surabhi Sachdev and Sarah Caudill and Heather Fong and Rico K. L. Lo and Cody Messick and Debnandini Mukherjee and Ryan Magee and Leo Tsukada and Kent Blackburn and Patrick Brady and Patrick Brockill and Kipp Cannon and Sydney J. Chamberlin and Deep Chatterjee and Jolien D. E. Creighton and Patrick Godwin and Anuradha Gupta and Chad Hanna and Shasvath Kapadia and Ryan N. Lang and Tjonnie G. F. Li and Duncan Meacher and Alexander Pace and Stephen Privitera and Laleh Sadeghian and Leslie Wade and Madeline Wade and Alan Weinstein and Sophia Liting Xiao},
      year={2019},
      eprint={1901.08580},
      archivePrefix={arXiv},
      primaryClass={gr-qc}
}

@article{Aubin:2020goo,
    author = "Aubin, F. and others",
    title = "{The MBTA pipeline for detecting compact binary coalescences in the third LIGO\textendash{}Virgo observing run}",
    eprint = "2012.11512",
    archivePrefix = "arXiv",
    primaryClass = "gr-qc",
    doi = "10.1088/1361-6382/abe913",
    journal = "Class. Quant. Grav.",
    volume = "38",
    number = "9",
    pages = "095004",
    year = "2021"
}

@misc{chu2021spiir,
      title={The SPIIR online coherent pipeline to search for gravitational waves from compact binary coalescences}, 
      author={Qi Chu and Manoj Kovalam and Linqing Wen and Teresa Slaven-Blair and Joel Bosveld and Yanbei Chen and Patrick Clearwater and Alex Codoreanu and Zhihui Du and Xiangyu Guo and Xiaoyang Guo and Kyungmin Kim and Tjonnie G. F. Li and Victor Oloworaran and Fiona Panther and Jade Powell and Anand S. Sengupta and Karl Wette and Xingjiang Zhu},
      year={2021},
      eprint={2011.06787},
      archivePrefix={arXiv},
      primaryClass={gr-qc}
}

@ARTICLE{2021SoftX..1400678D,
       author = {{Drago}, Marco and {Klimenko}, Sergey and {Lazzaro}, Claudia and {Milotti}, Edoardo and {Mitselmakher}, Guenakh and {Necula}, Valentin and {O'Brian}, Brendan and {Prodi}, Giovanni Andrea and {Salemi}, Francesco and {Szczepanczyk}, Marek and {Tiwari}, Shubhanshu and {Tiwari}, Vaibhav and {Gayathri}, V. and {Vedovato}, Gabriele and {Yakushin}, Igor},
        title = "{coherent WaveBurst, a pipeline for unmodeled gravitational-wave data analysis}",
      journal = {SoftwareX},
         year = 2021,
        month = jun,
       volume = {14},
          eid = {100678},
        pages = {100678},
          doi = {10.1016/j.softx.2021.100678},
archivePrefix = {arXiv},
       eprint = {2006.12604},
 primaryClass = {gr-qc},
       adsurl = {https://ui.adsabs.harvard.edu/abs/2021SoftX..1400678D}
}

@ARTICLE{2021ApJ...923..254D,
       author = {{Dal Canton}, Tito and {Nitz}, Alexander H. and {Gadre}, Bhooshan and {Cabourn Davies}, Gareth S. and {Villa-Ortega}, Ver{\'o}nica and {Dent}, Thomas and {Harry}, Ian and {Xiao}, Liting},
        title = "{Real-time Search for Compact Binary Mergers in Advanced LIGO and Virgo's Third Observing Run Using PyCBC Live}",
      journal = {\apj},
         year = 2021,
        month = dec,
       volume = {923},
       number = {2},
          eid = {254},
        pages = {254},
          doi = {10.3847/1538-4357/ac2f9a},
archivePrefix = {arXiv},
       eprint = {2008.07494},
 primaryClass = {astro-ph.HE},
       adsurl = {https://ui.adsabs.harvard.edu/abs/2021ApJ...923..254D}
}

@article{Cabero:2020eik,
    author = "Cabero, Miriam and Mahabal, Ashish and McIver, Jess",
    title = "{GWSkyNet: a real-time classifier for public gravitational-wave candidates}",
    eprint = "2010.11829",
    archivePrefix = "arXiv",
    primaryClass = "gr-qc",
    doi = "10.3847/2041-8213/abc5b5",
    journal = "Astrophys. J. Lett.",
    volume = "904",
    number = "1",
    pages = "L9",
    year = "2020"
}

@unpublished{jarov2023,
    author = "Seraphim Jarov and Sarah Thiele and Siddharth Soni and Julian Ding and
Jess McIver and Raymond Ng and Rikako Hatoya and Derek Davis",
    year = "in preparation",
}

@article{Heinzel:2023vkq,
    author = "Heinzel, Jack and Talbot, Colm and Ashton, Gregory and Vitale, Salvatore",
    title = "{Inferring the Astrophysical Population of Gravitational Wave Sources in the Presence of Noise Transients}",
    eprint = "2304.02665",
    archivePrefix = "arXiv",
    primaryClass = "astro-ph.HE",
    month = "4",
    year = "2023"
}

@article{Richards,
	doi = {10.1088/0004-637x/733/1/10},
  
	url = {https://doi.org/10.1088%2F0004-637x%2F733%2F1%2F10},
  
	year = 2011,
	month = {apr},
  
	publisher = {American Astronomical Society},
  
	volume = {733},
  
	number = {1},
  
	pages = {10},
  
	author = {Joseph W. Richards and Dan L. Starr and Nathaniel R. Butler and Joshua S. Bloom and John M. Brewer and Arien Crellin-Quick and Justin Higgins and Rachel Kennedy and Maxime Rischard},
  
	title = {{ON} {MACHINE}-{LEARNED} {CLASSIFICATION} {OF} {VARIABLE} {STARS} {WITH} {SPARSE} {AND} {NOISY} {TIME}-{SERIES} {DATA}
},
  
	journal = {The Astrophysical Journal}
}

@article{staccato,
    author = {Revsbech, E. A. and Trotta, R. and van Dyk, D. A.},
    title = "{STACCATO: a novel solution to supernova photometric classification with biased training sets}",
    journal = {Monthly Notices of the Royal Astronomical Society},
    volume = {473},
    number = {3},
    pages = {3969-3986},
    year = {2017},
    month = {10},
    issn = {0035-8711},
    doi = {10.1093/mnras/stx2570},
    url = {https://doi.org/10.1093/mnras/stx2570},
    eprint = {https://academic.oup.com/mnras/article-pdf/473/3/3969/21841665/stx2570.pdf},
}

@article{RFquasars,
	author = {{Carrasco, D.} and {Barrientos, L. F.} and {Pichara, K.} and {Anguita, T.} and {Murphy, D. N. A.} and {Gilbank, D. G.} and {Gladders, M. D.} and {Yee, H. K. C.} and {Hsieh, B. C.} and {L\'opez, S.}},
	title = {Photometric classification of quasars from RCS-2  using Random Forest},
	DOI= "10.1051/0004-6361/201525752",
	url= "https://doi.org/10.1051/0004-6361/201525752",
	journal = {A\&A},
	year = 2015,
	volume = 584,
	pages = "A44",
	month = "",
}

@article{sunspots,
author = {Stenning, David C. and Lee, Thomas C. M. and van Dyk, David A. and Kashyap, Vinay and Sandell, Julia and Young, C. Alex},
title = {Morphological feature extraction for statistical learning with applications to solar image data},
journal = {Statistical Analysis and Data Mining: The ASA Data Science Journal},
volume = {6},
number = {4},
pages = {329-345},
doi = {https://doi.org/10.1002/sam.11200},
url = {https://onlinelibrary.wiley.com/doi/abs/10.1002/sam.11200},
eprint = {https://onlinelibrary.wiley.com/doi/pdf/10.1002/sam.11200},
year = {2013}
}

@article{Biswas:2019wmx,
    author = "Biswas, Ayon and McIver, Jess and Mahabal, Ashish",
    title = "{New methods to assess and improve LIGO detector duty cycle}",
    eprint = "1910.12143",
    archivePrefix = "arXiv",
    primaryClass = "astro-ph.IM",
    doi = "10.1088/1361-6382/ab8650",
    journal = "Class. Quant. Grav.",
    volume = "37",
    number = "17",
    pages = "175008",
    year = "2020"
}

@article{Baker:2014eba,
    author = "Baker, Paul T. and Caudill, Sarah and Hodge, Kari A. and Talukder, Dipongkar and Capano, Collin and Cornish, Neil J.",
    title = "{Multivariate Classification with Random Forests for Gravitational Wave Searches of Black Hole Binary Coalescence}",
    eprint = "1412.6479",
    archivePrefix = "arXiv",
    primaryClass = "gr-qc",
    reportNumber = "LIGO-DOCUMENT-P1400231",
    doi = "10.1103/PhysRevD.91.062004",
    journal = "Phys. Rev. D",
    volume = "91",
    number = "6",
    pages = "062004",
    year = "2015"
}

@INPROCEEDINGS{5171374,
  author={Sfiligoi, Igor and Bradley, Daniel C. and Holzman, Burt and Mhashilkar, Parag and Padhi, Sanjay and Wurthwein, Frank},
  booktitle={2009 WRI World Congress on Computer Science and Information Engineering}, 
  title={The Pilot Way to Grid Resources Using glideinWMS}, 
  year={2009},
  volume={2},
  number={},
  pages={428-432},
  doi={10.1109/CSIE.2009.950}}

@article{Pordes:2007zzb,
    author = "Pordes, Ruth and others",
    editor = "Keyes, David E.",
    title = "{The Open Science Grid}",
    reportNumber = "FERMILAB-CONF-07-217-CD",
    doi = "10.1088/1742-6596/78/1/012057",
    journal = "J. Phys. Conf. Ser.",
    volume = "78",
    pages = "012057",
    year = "2007"
}

@ARTICLE{2020SciPy-NMeth,
  author  = {Virtanen, Pauli and Gommers, Ralf and Oliphant, Travis E. and
            Haberland, Matt and Reddy, Tyler and Cournapeau, David and
            Burovski, Evgeni and Peterson, Pearu and Weckesser, Warren and
            Bright, Jonathan and {van der Walt}, St{\'e}fan J. and
            Brett, Matthew and Wilson, Joshua and Millman, K. Jarrod and
            Mayorov, Nikolay and Nelson, Andrew R. J. and Jones, Eric and
            Kern, Robert and Larson, Eric and Carey, C J and
            Polat, {\.I}lhan and Feng, Yu and Moore, Eric W. and
            {VanderPlas}, Jake and Laxalde, Denis and Perktold, Josef and
            Cimrman, Robert and Henriksen, Ian and Quintero, E. A. and
            Harris, Charles R. and Archibald, Anne M. and
            Ribeiro, Ant{\^o}nio H. and Pedregosa, Fabian and
            {van Mulbregt}, Paul and {SciPy 1.0 Contributors}},
  title   = {{{SciPy} 1.0: Fundamental Algorithms for Scientific
            Computing in Python}},
  journal = {Nature Methods},
  year    = {2020},
  volume  = {17},
  pages   = {261--272},
  adsurl  = {https://rdcu.be/b08Wh},
  doi     = {10.1038/s41592-019-0686-2},
}

@Article{         harris2020array,
 title         = {Array programming with {NumPy}},
 author        = {Charles R. Harris and K. Jarrod Millman and St{\'{e}}fan J.
                 van der Walt and Ralf Gommers and Pauli Virtanen and David
                 Cournapeau and Eric Wieser and Julian Taylor and Sebastian
                 Berg and Nathaniel J. Smith and Robert Kern and Matti Picus
                 and Stephan Hoyer and Marten H. van Kerkwijk and Matthew
                 Brett and Allan Haldane and Jaime Fern{\'{a}}ndez del
                 R{\'{i}}o and Mark Wiebe and Pearu Peterson and Pierre
                 G{\'{e}}rard-Marchant and Kevin Sheppard and Tyler Reddy and
                 Warren Weckesser and Hameer Abbasi and Christoph Gohlke and
                 Travis E. Oliphant},
 year          = {2020},
 month         = sep,
 journal       = {Nature},
 volume        = {585},
 number        = {7825},
 pages         = {357--362},
 doi           = {10.1038/s41586-020-2649-2},
 publisher     = {Springer Science and Business Media {LLC}},
 url           = {https://doi.org/10.1038/s41586-020-2649-2}
}

@article{gwpy,
    title = "{GWpy: A Python package for gravitational-wave astrophysics}",
   author = {{Macleod}, D.~M. and {Areeda}, J.~S. and {Coughlin}, S.~B. and {Massinger}, T.~J. and {Urban}, A.~L.},
  journal = {SoftwareX},
   volume = 13,
    pages = 100657,
     year = 2021,
     issn = {2352-7110},
      doi = {10.1016/j.softx.2021.100657},
      url = {https://www.sciencedirect.com/science/article/pii/S2352711021000029},
}

@article{Nitz:2017svb,
    author = "Nitz, Alexander H. and Dent, Thomas and Dal Canton, Tito and Fairhurst, Stephen and Brown, Duncan A.",
    title = "{Detecting binary compact-object mergers with gravitational waves: Understanding and Improving the sensitivity of the PyCBC search}",
    eprint = "1705.01513",
    archivePrefix = "arXiv",
    primaryClass = "gr-qc",
    reportNumber = "LIGO-P1700088-V4",
    doi = "10.3847/1538-4357/aa8f50",
    journal = "Astrophys. J.",
    volume = "849",
    number = "2",
    pages = "118",
    year = "2017"
}

@article{Alvarez-Lopez:2023dmv,
    author = "Alvarez-Lopez, Sofia and Liyanage, Annudesh and Ding, Julian and Ng, Raymond and McIver, Jess",
    title = "{GSpyNetTree: A signal-vs-glitch classifier for gravitational-wave event candidates}",
    eprint = "2304.09977",
    archivePrefix = "arXiv",
    primaryClass = "gr-qc",
    month = "4",
    year = "2023"
}

@article{Glanzer:2022avx,
    author = "Glanzer, J. and others",
    title = "{Data quality up to the third observing run of advanced LIGO: Gravity Spy glitch classifications}",
    eprint = "2208.12849",
    archivePrefix = "arXiv",
    primaryClass = "gr-qc",
    reportNumber = "LIGO-P2200238",
    doi = "10.1088/1361-6382/acb633",
    journal = "Class. Quant. Grav.",
    volume = "40",
    number = "6",
    pages = "065004",
    year = "2023"
}

@article{Soni:2021cjy,
    author = "Soni, S. and others",
    title = "{Discovering features in gravitational-wave data through detector characterization, citizen science and machine learning}",
    eprint = "2103.12104",
    archivePrefix = "arXiv",
    primaryClass = "gr-qc",
    doi = "10.1088/1361-6382/ac1ccb",
    journal = "Class. Quant. Grav.",
    volume = "38",
    number = "19",
    pages = "195016",
    year = "2021"
}

@ARTICLE{2019MNRAS.484.4008G,
       author = {{Gaebel}, Sebastian M. and {Veitch}, John and {Dent}, Thomas and {Farr}, Will M.},
        title = "{Digging the population of compact binary mergers out of the noise}",
      journal = {\mnras},
         year = 2019,
        month = apr,
       volume = {484},
       number = {3},
        pages = {4008-4023},
          doi = {10.1093/mnras/stz225},
archivePrefix = {arXiv},
       eprint = {1809.03815},
 primaryClass = {astro-ph.IM},
       adsurl = {https://ui.adsabs.harvard.edu/abs/2019MNRAS.484.4008G}
}

@article{PhysRevD.102.083026,
  title = {Gravitational-wave inference in the catalog era: Evolving priors and marginal events},
  author = {Galaudage, Shanika and Talbot, Colm and Thrane, Eric},
  journal = {Phys. Rev. D},
  volume = {102},
  issue = {8},
  pages = {083026},
  numpages = {24},
  year = {2020},
  month = {Oct},
  publisher = {American Physical Society},
  doi = {10.1103/PhysRevD.102.083026},
  url = {https://link.aps.org/doi/10.1103/PhysRevD.102.083026}
}

@article{PhysRevD.102.123022,
  title = {Binary black hole mergers from LIGO/Virgo O1 and O2: Population inference combining confident and marginal events},
  author = {Roulet, Javier and Venumadhav, Tejaswi and Zackay, Barak and Dai, Liang and Zaldarriaga, Matias},
  journal = {Phys. Rev. D},
  volume = {102},
  issue = {12},
  pages = {123022},
  numpages = {24},
  year = {2020},
  month = {Dec},
  publisher = {American Physical Society},
  doi = {10.1103/PhysRevD.102.123022},
  url = {https://link.aps.org/doi/10.1103/PhysRevD.102.123022}
}

@article{LIGOScientific:2020iuh,
    author = "Abbott, R. and others",
    collaboration = "LIGO Scientific, Virgo",
    title = "{GW190521: A Binary Black Hole Merger with a Total Mass of $150  M_{\odot}$}",
    eprint = "2009.01075",
    archivePrefix = "arXiv",
    primaryClass = "gr-qc",
    doi = "10.1103/PhysRevLett.125.101102",
    journal = "Phys. Rev. Lett.",
    volume = "125",
    number = "10",
    pages = "101102",
    year = "2020"
}

@article{Christensen:2022bxb,
    author = "Christensen, Nelson and Meyer, Renate",
    title = "{Parameter estimation with gravitational waves}",
    eprint = "2204.04449",
    archivePrefix = "arXiv",
    primaryClass = "gr-qc",
    doi = "10.1103/RevModPhys.94.025001",
    journal = "Rev. Mod. Phys.",
    volume = "94",
    number = "2",
    pages = "025001",
    year = "2022"
}

@article{Callister:2017urp,
    author = "Callister, T. A. and Kanner, J. B. and Massinger, T. J. and Dhurandhar, S. and Weinstein, A. J.",
    title = "{Observing Gravitational Waves with a Single Detector}",
    eprint = "1704.00818",
    archivePrefix = "arXiv",
    primaryClass = "astro-ph.IM",
    doi = "10.1088/1361-6382/aa7a76",
    journal = "Class. Quant. Grav.",
    volume = "34",
    number = "15",
    pages = "155007",
    year = "2017"
}

@article{Nitz:2020naa,
    author = "Nitz, Alexander H. and Dent, Thomas and Davies, Gareth S. and Harry, Ian",
    title = "{A Search for Gravitational Waves from Binary Mergers with a Single Observatory}",
    eprint = "2004.10015",
    archivePrefix = "arXiv",
    primaryClass = "astro-ph.HE",
    doi = "10.3847/1538-4357/ab96c7",
    journal = "Astrophys. J.",
    volume = "897",
    number = "2",
    pages = "169",
    month = "4",
    year = "2020"
}

@article{Davies:2022thw,
    author = "Davies, Gareth S. Cabourn and Harry, Ian W.",
    title = "{Establishing significance of gravitational-wave signals from a single observatory in the PyCBC offline search}",
    eprint = "2203.08545",
    archivePrefix = "arXiv",
    primaryClass = "gr-qc",
    doi = "10.1088/1361-6382/ac8862",
    journal = "Class. Quant. Grav.",
    volume = "39",
    number = "21",
    pages = "215012",
    year = "2022"
}

@book{whittle1951hypothesis,
  title={Hypothesis Testing in Time Series Analysis},
  author={Whittle, P.},
  isbn={9780598919823},
  lccn={52021616},
  series={Statistics / Uppsala universitet},
  url={https://books.google.co.in/books?id=nE\_QAAAAMAAJ},
  year={1951},
  publisher={Almqvist \& Wiksells boktr.}
}

@article{Cabero:2019orq,
    author = "Cabero, Miriam and others",
    title = "{Blip glitches in Advanced LIGO data}",
    eprint = "1901.05093",
    archivePrefix = "arXiv",
    primaryClass = "physics.ins-det",
    doi = "10.1088/1361-6382/ab2e14",
    journal = "Class. Quant. Grav.",
    volume = "36",
    number = "15",
    pages = "15",
    year = "2019"
}

@article{PhysRevD.107.024030,
  title = {Deep learning network to distinguish binary black hole signals from short-duration noise transients},
  author = {Choudhary, Sunil and More, Anupreeta and Suyamprakasam, Sudhagar and Bose, Sukanta},
  journal = {Phys. Rev. D},
  volume = {107},
  issue = {2},
  pages = {024030},
  numpages = {12},
  year = {2023},
  month = {Jan},
  publisher = {American Physical Society},
  doi = {10.1103/PhysRevD.107.024030},
  url = {https://link.aps.org/doi/10.1103/PhysRevD.107.024030}
}

@article{Kapadia:2017fhb,
    author = "Kapadia, Shasvath J. and Dent, Thomas and Dal Canton, Tito",
    title = "{Classifier for gravitational-wave inspiral signals in nonideal single-detector data}",
    eprint = "1709.02421",
    archivePrefix = "arXiv",
    primaryClass = "astro-ph.IM",
    doi = "10.1103/PhysRevD.96.104015",
    journal = "Phys. Rev. D",
    volume = "96",
    number = "10",
    pages = "104015",
    year = "2017"
}

@article{Pratten:2020ruz,
    author = "Pratten, Geraint and Vecchio, Alberto",
    title = "{Assessing gravitational-wave binary black hole candidates with Bayesian odds}",
    eprint = "2008.00509",
    archivePrefix = "arXiv",
    primaryClass = "gr-qc",
    doi = "10.1103/PhysRevD.104.124039",
    journal = "Phys. Rev. D",
    volume = "104",
    number = "12",
    pages = "124039",
    year = "2021"
}

\end{document}